\documentclass[12pt]{article}
\usepackage{amsfonts,epsfig,latexsym,amscd}
\usepackage{amssymb}%,letterspace}
\usepackage{amsmath}
\usepackage{amssymb}
\usepackage{mathrsfs}
\usepackage{bbm}
\usepackage[hidelinks]{hyperref}
\usepackage{graphicx}

\newcommand{\R}{{\mathbb{R}}}

\newcommand{\be}{\begin{equation}}
\newcommand{\ee}{\end{equation}}
\newcommand{\bea}{\begin{eqnarray}}
\newcommand{\eea}{\end{eqnarray}}
\newcommand{\bean}{\begin{eqnarray*}}

\newcommand{\eean}{\end{eqnarray*}}
\font\upright=cmu10 scaled\magstep1

\newcommand{\PP}{\hbox{\upright\rlap{I}\kern 1.5pt P}}

\newcommand{\identity}{{\upright\rlap{1}\kern 2.0pt 1}}

\newcommand{\HH}{\mbox{\hbox{\upright\rlap{I}\kern 1.7pt H}}}

\newcommand{\fr}{\frac}
\newcommand{\lm}{\lambda}
\newcommand{\ra}{\rightarrow}
\newcommand{\al}{\alpha}

\newcommand{\acc}{\\[3mm]}

\usepackage{subfigure}

\def\d{\mathrm{d}}
\def\e{\mathrm{e}}

\def\kihagy#1{}

\def\Tr{\ensuremath{\mathop{\rm tr}}}

\begin{document}
\begin{center}
{ \Large \bf  Time-Dependent BPS Skyrmions}\acc
{\large Theodora Ioannidou\footnote{{\it Email}: {\sf ti3@auth.gr}} and 
\'Arp\'ad Luk\'acs\footnote{{\it Email}:  
{\sf arpadlukacs@civil.auth.gr }}} \acc
\noindent{\em Faculty of Civil Engineering,  School of Engineering, 
Aristotle University of Thessaloniki, Thessaloniki 54124, Greece} 
\end{center}

\begin{abstract}
An extended version  of the BPS Skyrme model  that admits  time-dependent solutions is discussed. Initially, by introducing a power law at the original potential term of the BPS Skyrme model  the existence, stability and structure of the corresponding solutions is investigated.
 Then, the frequencies and half-lifes of the radial oscillations of   the constructed time-dependent solutions  are  determined.
\end{abstract}
 
%%%%%%%%%%%%%%%%%%%%%%%%%
%%%%%%%%%%%%%%%%%%%%%%%%%%
\section{Introduction}

The classical Skyrme model \cite{Skyrme} is a strong candidate for describing  the low energy regime of quantum chromodynamics (for up-to-date research, see for example, Ref. \cite{Manton}).
The Skyrme model is  a nonlinear theory of mesons in three space dimensions admitting as solutions topological solitons, the so-called skyrmions. 
Each skyrmion is associated with a topological charge, interpreted as baryon number $n$; and therefore, skyrmions are baryons emerging from a meson field.
Since its original formulation, the Skyrme model has been able to predict the nucleon properties up to thirty precent. 
Therefore, several modifications of the classical model have been considered aiming to improve the predictions.

One such modification was introduced in \cite{Jac} by adding a sextic term (instead of  a quartic one)  in the classical Skyrme model Lagrangian,  for stabilising the skyrmions. 
Alternatively, both terms can be retained.
However, the latter can produce some problems since   in the {\it non static case} the two time derivatives  lead to pathological runaway solutions. 
This is not true  in the {\it static case} as shown in \cite{Piette}, where the classical properties of a generalised Skyrme model  by considering a combination of  the quartic and  sextic term  have been studied. 
 The  corresponding  skyrmions were obtained following numerical  techniques and approximations used  in the classical case and it was shown that they possess the same symmetries as the classical ones but are more bound.
 
 Recently in \cite{Adametal}, a submodel of the generalised Skyrme model has been considered, which consists, {\it only},  of the square of the baryon current and a potential term.  This model is called  the BPS Skyrme model  since a  Bogomolny bound exists and the static solution saturates it.
By choosing a particular type of the potential,  the BPS Skyrme model admits  topological compactons (solitons that reach the exact value at a finite distance; that is,  with compact support) which can be obtained analytically and  reproduce  features and properties of the liquid drop model of nuclei (for details see, for example, Ref. \cite{Ad} and references in it). 
However, most of the investigations have  considered the static case since the BPS Skyrme model does not have a well defined Cauchy problem due to  the non-standard kinetic term and the non-analytic behaviour of the compactons at the boundaries.

In this paper we  consider  the dynamics of the BPS Skyrme model and investigate the existence, stability and structure of time-dependent configurations. Initially,   a generalised version of the potential is introduced.  Then, the existence  of the corresponding solitons is studied for different values of the parameter. In addition, it is shown how by varying the parameter the BPS skyrmions transform from solitons to compactons. 

%%%%%%%%%%%%%%%%%%%%%%
%%%%%%%%%%%%%%%%%%%%%%%
\section{The BPS Skyrme Model}

The action of the BPS Skyrme model is defined by
\begin{equation}
  \label{eq:lag}
  S = \int \d^4 x \left\{
    -\lambda^2 \pi^2 B^\mu B_\mu -\mu^2\, V(U,U^\dag)  \right\},
\end{equation}
 where $U(t,{\bf x})$ is the Skyrme field (that is, an $SU(2)$-valued scalar field);   $\mu$ is a free parameter with units  MeV$^2$;  $V(U,U^\dag)$ is the potential (or mass)  term which breaks the chiral symmetry of the model; $\lm$ is a positive constant with units MeV$^{-1}$;
and $B^\mu$ is the topological current density  defined by
\begin{equation}
  \label{eq:topcurr}
  B^\mu = \frac{1}{24\pi^2}\,\epsilon^{\mu\nu\rho\sigma}\Tr\left(L_\nu L_\rho L_\sigma\right),
\end{equation}
where $L_\mu = U^\dagger \partial_\mu U$ is the $su(2)$-valued current and  $g_{\mu\nu}={\rm diag}(1,-1,-1,-1)$ is the Minkowski metric.
Using scaling arguments it can be shown that the first  term prevents the BPS skyrmions from shrinking to zero size while the potential term stabilizes them against arbitrary expansion.  
By construction, the BPS model is more topological in nature than the classical Skyrme model.

By rescaling  $x^\mu \to \left(\lambda n / \sqrt{2}\pi\mu\right)^{1/3}x^\mu$ where $n$ being the baryon number, the action \ (\ref{eq:lag}) becomes
\begin{equation}
  \label{mm}
  S =- \frac{\mu^2}{2}\left( \frac{\lambda n}{\sqrt{2}\pi\mu}\right)^{4/3} \int \d^4 x \left\{
    \frac{1}{144n^2} \left[ \epsilon^{\mu\nu\rho\sigma}\Tr(L_\nu L_\rho L_\sigma)\right]^2 +2 V(U,U^\dag)\right\}.
\end{equation}
In what follows, we consider  the action  (\ref{mm}) divided  by   $\left(\lambda n/\sqrt{2}\pi\mu\right)^{4/3}$.

Similarly to the classical case \cite{MI}, we parametrize $U$ by a real scalar field $f$ and a three component unit vector $\hat{\bf n}$ as
\begin{equation}
  \label{eq:param}
  U = \exp\left( i  f\,  \vec{\bf \sigma}  \cdot \hat{\bf n} \right)
\end{equation}
where $\vec{\bf \sigma}=(\sigma^1,\sigma^2,\sigma^3)$ are the Pauli matrices.
The vector field is related to a complex scalar field $\psi$ by the stereographic projection
\begin{equation}
  \label{eq:param_n}
  \hat{\bf n} = \frac{1}{1+|\psi|^2}\left(\psi + \bar{\psi},-i(\psi-\bar{\psi}), 1-|\psi|^2\right).
\end{equation}

For simplicity, we consider spherical symmetry. 
This is done by separation of the radial and  angular dependence of the fields. 
In particular, using the polar coordinates $(r,\theta,\phi)\in\R^3$ we assume  that   $f=f(r,t)$ and  
$\psi=\psi(\vartheta,\varphi)\equiv \tan\left(\frac{\vartheta}{2}\right) \e^{in\varphi}$.
Then,  the action  (\ref{mm}) simplifies to 
\begin{equation}
  \label{eq:lag_rad}
  S= 2\pi\mu^2\int\d t\int \d r\, r^2 \left[ \frac{\sin^4 f}{r^4}\left( \dot{f}^2 - f'^2 \right) -2 V\right],
\end{equation}
and the corresponding equation of motion becomes
\begin{equation}
  \label{eq:radeq}
  \frac{\sin^4 f}{r^4}\left( \ddot{f} - f'' +\frac{2}{r}\,f' \right) + \frac{2\sin^3 f\cos f}{r^4}\left( \dot{f}^2 - f'^2 \right)
  +\frac{\partial V}{\partial f} =0.
\end{equation}

It can be easily observed that the form of the potential plays a critical role for the existence and   type of the corresponding solutions. Let the potential to be  given by the power law 
\bea
  \label{eq:pots}
  V_\al &=& \fr{1}{2}\left(2 - \Tr U \right)^\al\nonumber\acc
  &=&\left(1-\cos f\right)^\al,
\eea
where $\al\in \R$ is a free parameter. 
Then,  for $\al=1$ the BPS  Skyrme model studied in \cite{Adametal,Ad} is obtained.
 For $\al<3$, the solutions are compactons with a non-analytical behaviour at the boundaries. 
Finally, for $\al\ge 3$ time-dependent skyrmions can be derived as it is shown in Section \ref{sec:timedept}.
The aforementioned solutions are similar in their properties, since topology predominates  the type of the potential.

%%%%%%%%%%%%%%%%%%%%%%
%%%%%%%%%%%%%%%%%%%%
\section{Static BPS Skyrmions}

Let us concentrate on the  static case, i.e., $f=f(r)$. Then, the energy (\ref{eq:lag_rad})  can be expressed as a sum of a square and a topological quantity. That is,
\begin{equation}
  \label{eq:BPS_erg}
  E =  4\pi\int\d r \,r^2 \left[ \left( \frac{\sin^2 f}{r^2}f' \pm  \sqrt{2 V}\right)^2 \mp \frac{\sqrt{2V}\sin^2 f}{r^2}f'\right].
\end{equation}
Since the last term is topological invariant,  in each topological sector a minimum is obtained satisfying  the Bogomolny-Prasad-Sommerfield (BPS) equation
\begin{equation}
  \label{eq:BPS_eq}
  f' = \mp \frac{\sqrt{2V}r^2}{\sin^2 f}
\end{equation}
 i.e., when the square  term of (\ref{eq:BPS_erg}) vanishes. Solutions of equation (\ref{eq:BPS_eq}) can be obtained either analytically or numerically depending on the form of the potential. 
In what follows, we assume that the potential  is given by equation (\ref{eq:pots}) and investigate the type and form of the obtained solutions for different values of $\al$.

In the simplest case where  $\al=1$, a closed form for the profile function (\ref{eq:BPS_eq}) exists.
The obtained solutions (with different winding numbers) are related to each other since
by rescaling the coordinates the  winding number $n$ of the  ansatz  drops out of equation (\ref{eq:BPS_eq}).

Next, let us keep $\al$ arbitrary and study  the asymptotic behaviour of the solution (\ref{eq:BPS_eq}).
Since the potential vanishes for $f=2n\pi$ for $n\in\mathbb{Z}$, we consider $f(0)=\pi$ and $f(\infty)=0$. This choice of boundary conditions  sets   the baryon number equal to  the winding number. 

At spatial infinity, equation (\ref{eq:BPS_eq}) takes the form
\begin{equation}
  \label{eq:BPSradAsy}
  f^{2-\al} f' = \mp \frac{r^2}{2^{(\al-1)/2}},
\end{equation}
and its  solution (for $\al\ne 3$) is given by  
\begin{equation}
  \label{eq:phiAsy}
  f(r)^{3-\al} = \mp \fr{1}{2^{(\al-1)/2}}\left[\left(1-\fr{\al}{3}\right)r^3 + c\right].
\end{equation}

For $\al>3$, the solution has a power law tail since  $f(r)\propto r^\fr{1}{1-\al/3}$; while for  $\al=3$ it becomes exponentially localised since 
\begin{equation*}
  \label{eq:phiAsy3}
  f(r)\sim \exp\left( -\frac{r^3}{6}\right).
\end{equation*}

Finally, for $\al< 3$ a non-analytic behaviour at the boundary leads to  compacton solutions since for $r<r_b$, the  profile function (\ref{eq:BPS_eq}) is given by
\begin{equation*}
  \label{eq:compacton}
  f(r) \sim \left(r_b -r\right)^\gamma \left[ c_0 + c_1\left(r-r_b\right) + \dots\right],
\end{equation*}
where
\begin{equation*}
  \label{eq:compactonrb}
  \gamma=\frac{1}{3-\al},\quad \quad \ c_0 = \left[\sqrt{2}r_b^2 \left(3-\al\right)\right]^\gamma,\ \quad \quad c_1 = \sqrt{2}\,r_b\, c_0^{\al-2}.
\end{equation*}
Recall that, at the origin, the profile function of a regular solution is given by: $f = \pi + f'\left(0\right)r + \mathcal{O}(r^2)$.

\begin{figure}
%\vskip -25mm
\hfil\includegraphics[angle=0,scale=.5]{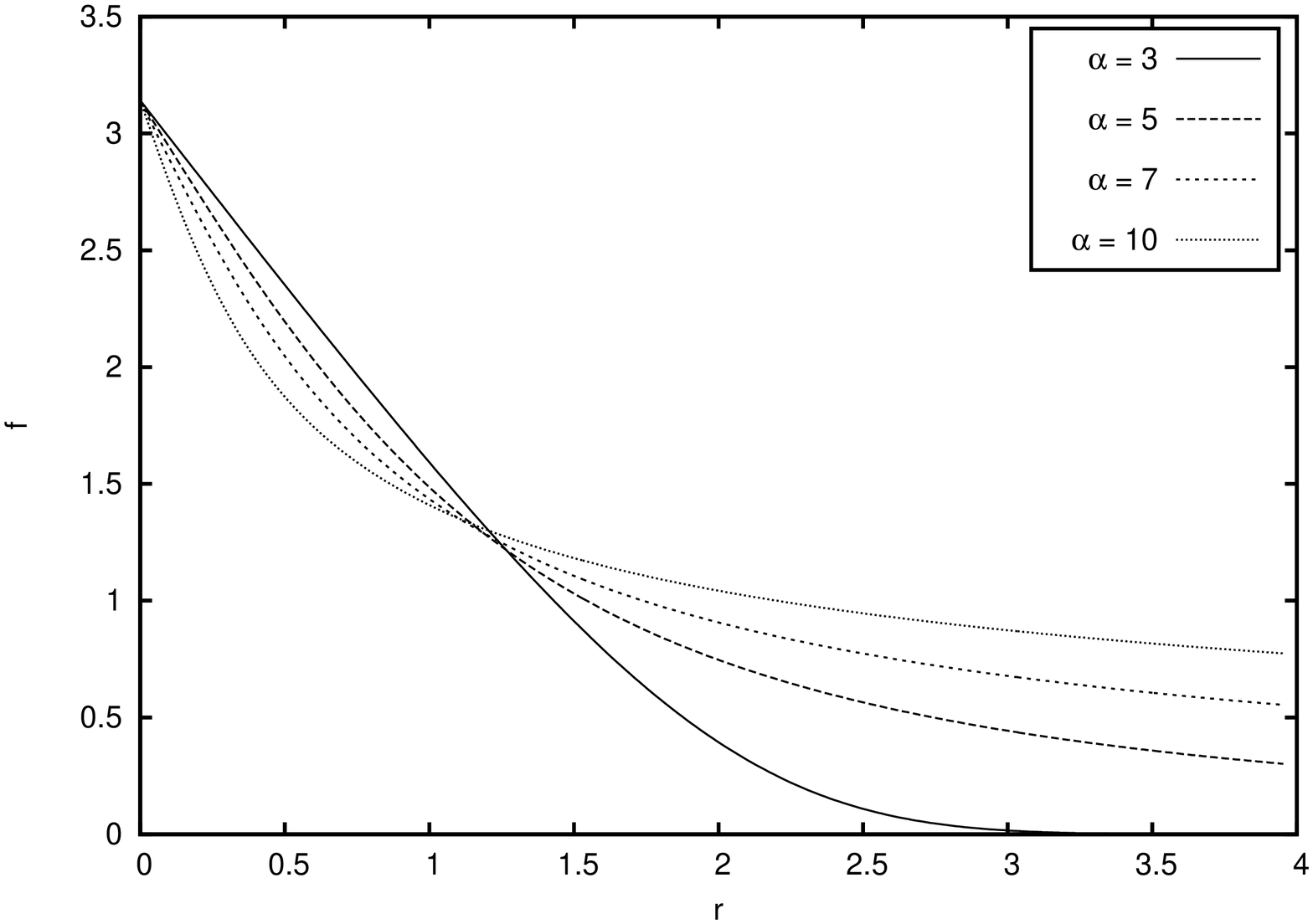}
%\vskip -13mm
\caption{The profile function  given by (\ref{eq:BPS_eq}) for different values of  $\al$.}
\label{fig:profs}
%\vskip 5mm
\end{figure}

\begin{figure}
\noindent\hfil\includegraphics[angle=0,scale=.5]{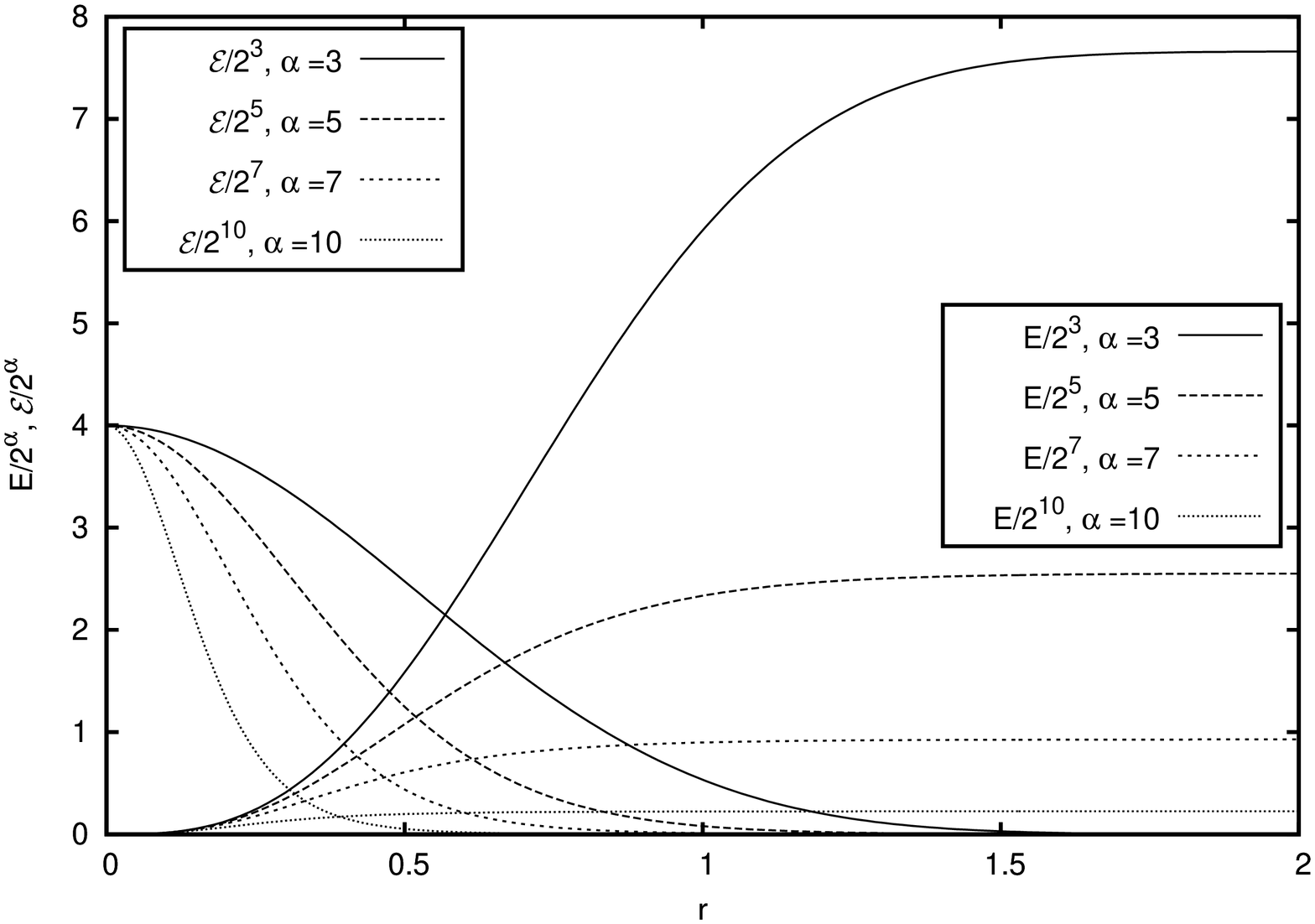}
%\vskip -13mm
\caption{The total energy and  energy density given by (\ref{eq:BPS_erg})  for different values of $\al$.}
\label{fig:ergs}
\end{figure}

\begin{figure}
%\vskip -15mm
\noindent\hfil\includegraphics[angle=0,scale=.5]{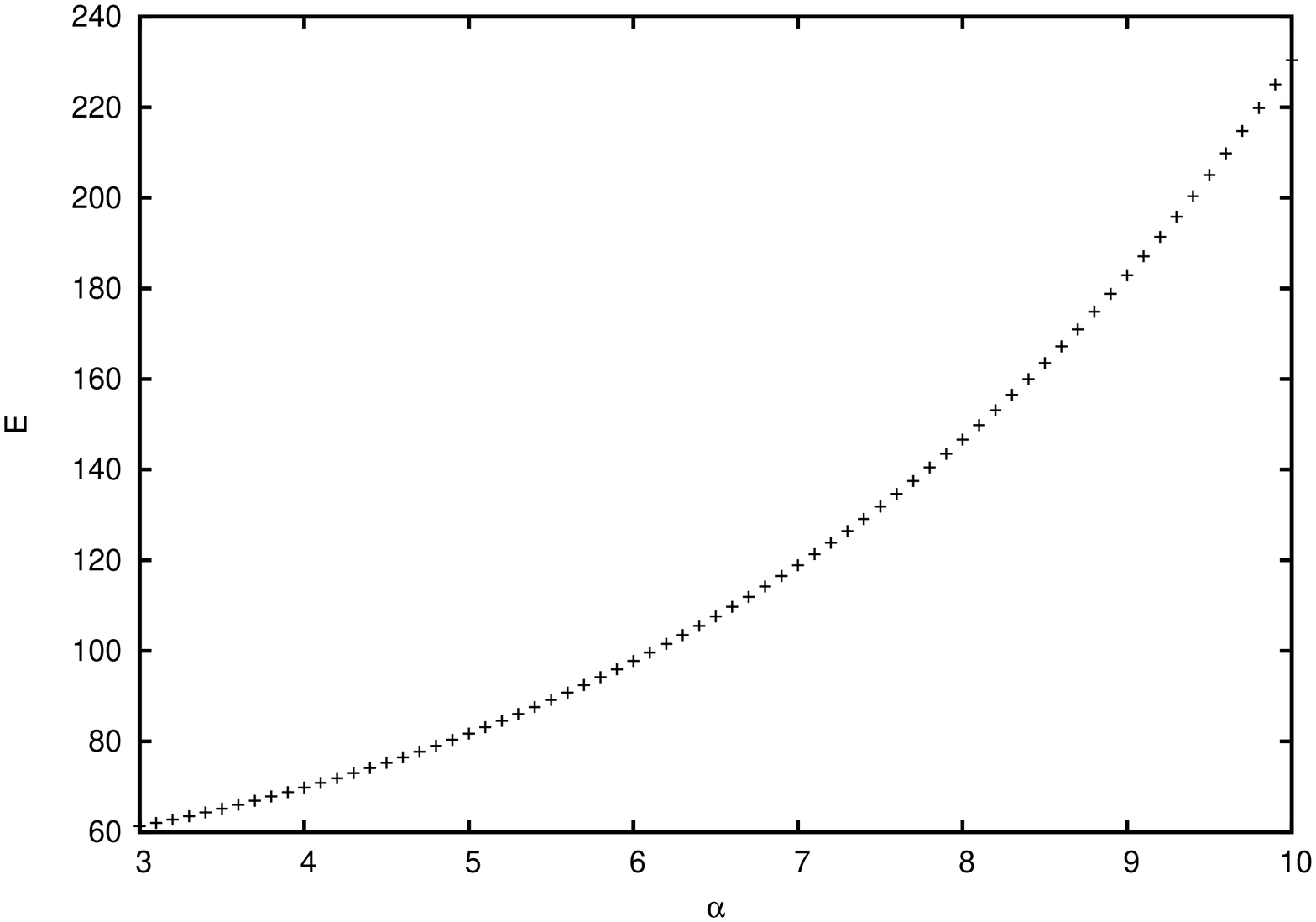}
%\vskip -13mm
\caption{The total energy (\ref{eq:BPS_erg}) as a function of $\al$.}
\label{fig:erga}
\end{figure}

The profile equation  (\ref{eq:BPS_eq}) of the aforementioned  cases have been obtained numerically  using   Dormand--Prince 8\textsuperscript{th} order
Runge--Kutta integration with embedded error estimation and adaptive step size control \cite{NR}. 
The results are presented in  Figure \ref{fig:profs} which displays the  solutions of (\ref{eq:BPS_eq}) (with the  minus sign) for different values of $\al\geq 3$.
In addition,  Figure \ref{fig:ergs} presents the total energy  (\ref{eq:BPS_erg}) and energy density $\mathcal{E}$ (i.e., $E(R)=4\pi \int_0^R  \d r\, r^2 \mathcal{E}$)   for different values of $\al$.
 Finally, Figure  \ref{fig:erga} displays the total energy (\ref{eq:BPS_erg}) as a function of $\al$.

%%%%%%%%%%%%%%%%%%%%%%%%%
%%%%%%%%%%%%%%%%%%%%%%%%%
\section{Dynamics of BPS Skyrmions}\label{sec:timedept}

Let us conclude by concentrating on  the non static case. 
For $\al>3$,   the model has a
well-defined dynamics since  the  principal part of the differential equation  (\ref{eq:radeq}) ,  $\ddot{f}-f''$, is  the same as that of the Klein-Gordon one.

To  study the radial excitations of the BPS skyrmions we describe  the soliton oscillations by  collective coordinates \cite{Rice, dyn}.
In particular, the radial pulsations are imposed due to the  variational ansatz
\begin{equation}
  \label{eq:rice}
  f(r,t) = f_0\left(\fr{r}{\rho(t)}\right),
\end{equation}
where $f_0$ is the static profile function by setting $r\ra r/\rho(t)$.  Then the action (\ref{eq:lag_rad}) takes the form
\begin{equation}
  \label{eq:lag_coll}
  L_\rho = K \,\frac{\dot{\rho}^2}{\rho^3} - U_1\,\frac{1}{\rho} + U_2\, \rho^3,
\end{equation}
where $K$, $U_1$, $U_2$ are defined by the integrals
\begin{equation}
  \label{eq:osc_coeff}
  \begin{aligned}
    K\, &= 4\pi\int \d r\, \sin^4 f_0\, f_0'^2 ,\\
    U_1 &=  4\pi\int \d r\, \frac{\sin^4 f_0}{r^2} \,f_0'^2 ,\\
    U_2 &= 8\pi \int \d r\,r^2\, V\left(f_0\right ).
  \end{aligned}
\end{equation}

\begin{table}
\begin{center}
\begin{tabular}{|c||c| c c c | c|}
\hline
$\al$   &  $E$    &  $K$   & $U_1$  & $U_2$   & $\omega_{\rm P}$ \\
\hline\hline
4       & 69.789  & 19.066  & 34.894  & 34.894  & 3.314 \\
5       & 81.711  & 17.632  & 40.850  & 40.850  & 3.728 \\
6       & 97.731  & 16.603  & 48.851  & 48.851  & 4.202 \\
7       & 118.881 & 15.863  & 59.417  & 59.417  & 4.741 \\
8       & 146.611 & 15.340  & 73.275  & 73.275  & 5.354 \\
9       & 182.892 & 14.987  & 91.411  & 91.411  & 6.049 \\
10      & 230.371 & 14.772  & 115.148 & 115.148 & 6.839 \\
\hline
\end{tabular}
\end{center}
\caption{Numerical data of the  static solution for different values of $\al$.}
\label{tab:stat}
\end{table}

For small oscillations,  $L_\rho$ needs to be expanded up to second order around $\rho=1$. Then the  first order term vanishes (that is,  $U_1=U_2$) since $f_0$ is a solution of the static field equations; while the second order term is equal to  the Lagrangian of a harmonic oscillator with frequency
\begin{equation}
  \label{eq:frq_coll}
  \omega_{\rm P}^2 = \frac{6\, U_2}{K}.
\end{equation}
The precision of the predicted frequencies deteriorates as $\al$ increases. 
That happens because as $\al$ increases  the tail of the oscillation becomes predominant and thus, the shape of the solution changes. In Table \ref{tab:stat}, the values of the 
 constants $K_1$, $U_1$, $U_2$ and the oscillation frequencies $\omega_{\rm P}$ are given for different values of $\al$. 
The vanishing of the first order term of the Lagrangian (that is,  $U_1=U_2$)  verifies the precision of our numerics. 

For the numerical simulations, equation (\ref{eq:radeq}) is discretised on a uniform grid in space and time. 
In particular, for the spatial coordinate, a  second order finite difference discretisation and for the time coordinate, the fourth order Runge-Kutta method (method of lines \cite{Schiesser}) have been considered. 
The default grid size is $r=[0\dots 20]$ and  $t=[0\dots 10]$ with $8\times 10^3$ spatial and $10^4$ temporal grid points (due to  Courant stability criterion). 
The accuracy of the results have been verified by  running simulations on a longer radial interval and half step sizes.

\begin{figure}
%\vskip 5mm
\noindent\hfil\includegraphics[angle=0,scale=.5]{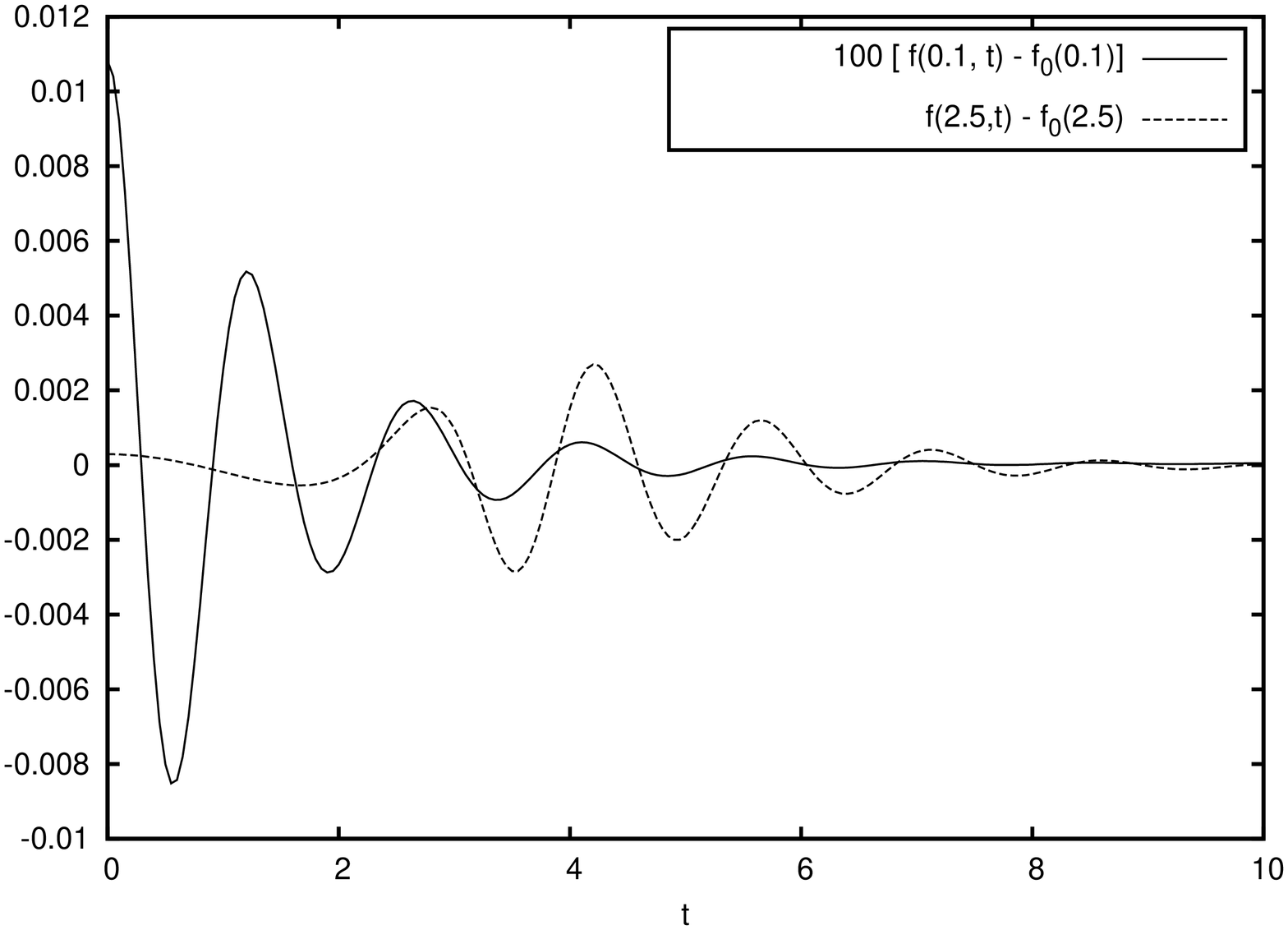}
%\vskip -13mm
\caption{Time evolution of an oscillating skyrmion with initial stretching $0.05 \%$  and $\al=6$, at two different points.}
\label{fig:short5}
\end{figure}

In addition, the  Neumann boundary conditions at large $r$ have been applied.
 As  initial data a uniformly stretched skyrmion has been used.
Then, the   frequencies and the decay constants of the oscillation modes have been measured; verifying that they are independent of  the amplitude of the stretch and  the parameters of the simulation.
For large values of  time  ($t>1$) and  given  $r$ (inside the soliton core, e.g., $r=0.05$), the decaying oscillations are approximated by $f(x,t)-f_0(x) \approx A_x \cos\left(\omega t+\delta\right) /2^{t/T}$.
Then, for the interval $t=1\dots 7$, the  amplitude $A_x$, the frequency $\omega$ and the half-life $T$ are obtained (with below $1\%$ precision) by a least squares fit.
In Table \ref{tab:dyn}, the frequencies and half-lifes of the deformed skyrmion oscillations
are presented and compared with the one  (that is, $\omega_{\rm P}$) obtained from the collective coordinate approximation. Note that, there is a difference between the measured frequencies and the ones obtained by the uniform oscillation
approximation \cite{dyn}. 
 This is  explained by studying the time evolution of the deformed skyrmions.  
 As demonstrated in Figure
\ref{fig:short5}, in the interior of the skyrmion  the deformation (described by the continues line) is a genuine pulsation while in the exterior    the deformation (described by the dashed line) is a wave packet travelling outwards. 

\begin{figure}
%\vskip -15mm
\noindent\hfil\includegraphics[angle=0,scale=.5]{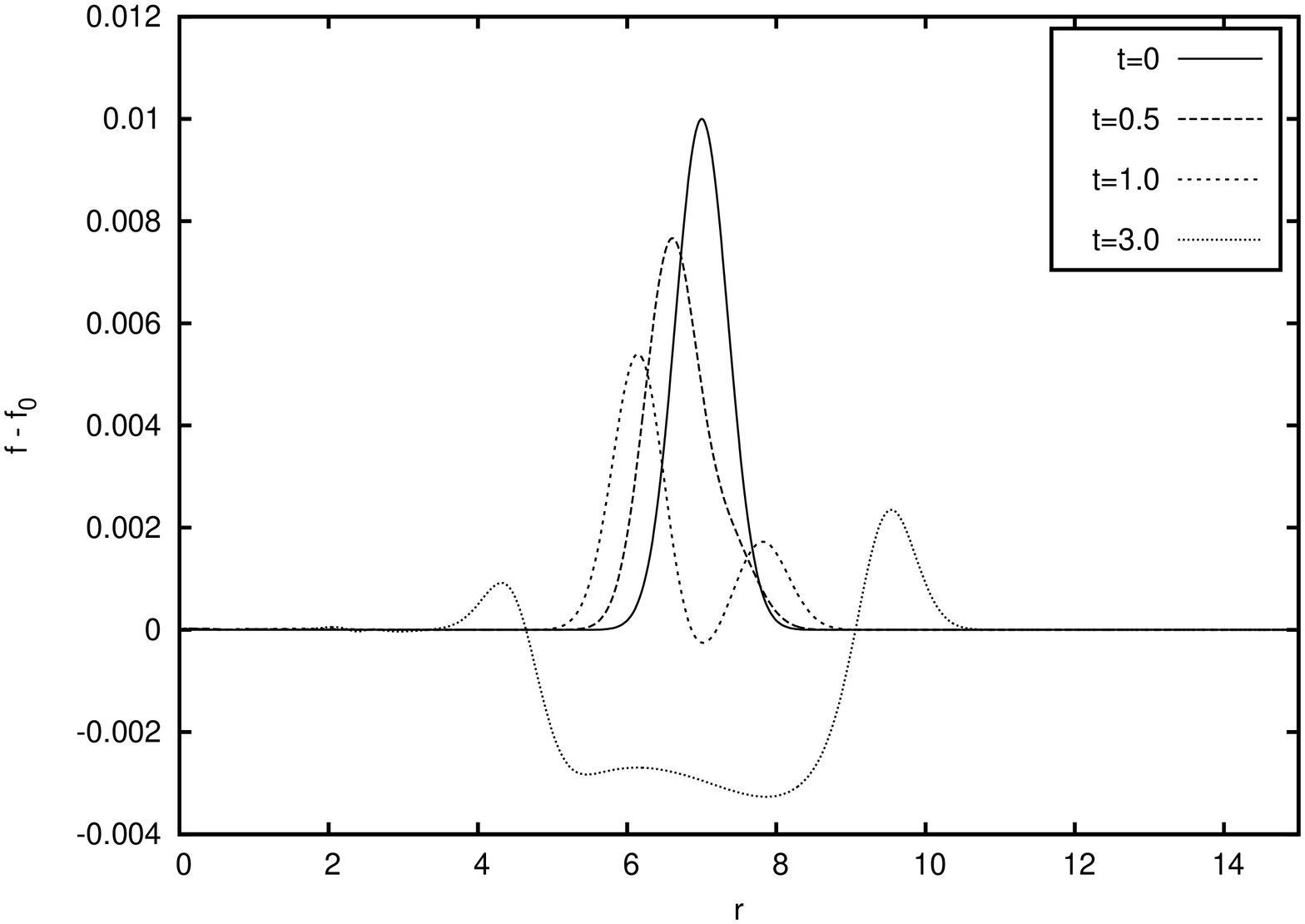}
%\vskip -13mm
\caption{Time evolution of a small perturbation on a skyrmion when $\al=6$. The amplitude of the initial Gaussian is $0.01$,
  its velocity  $-0.8$ (moving in), its squared width 0.25 and it is centered at $r=7$.}
\label{fig:wave1}
%\vskip 5mm
\end{figure}

\begin{figure}
\noindent\hfil\includegraphics[angle=0,scale=.5]{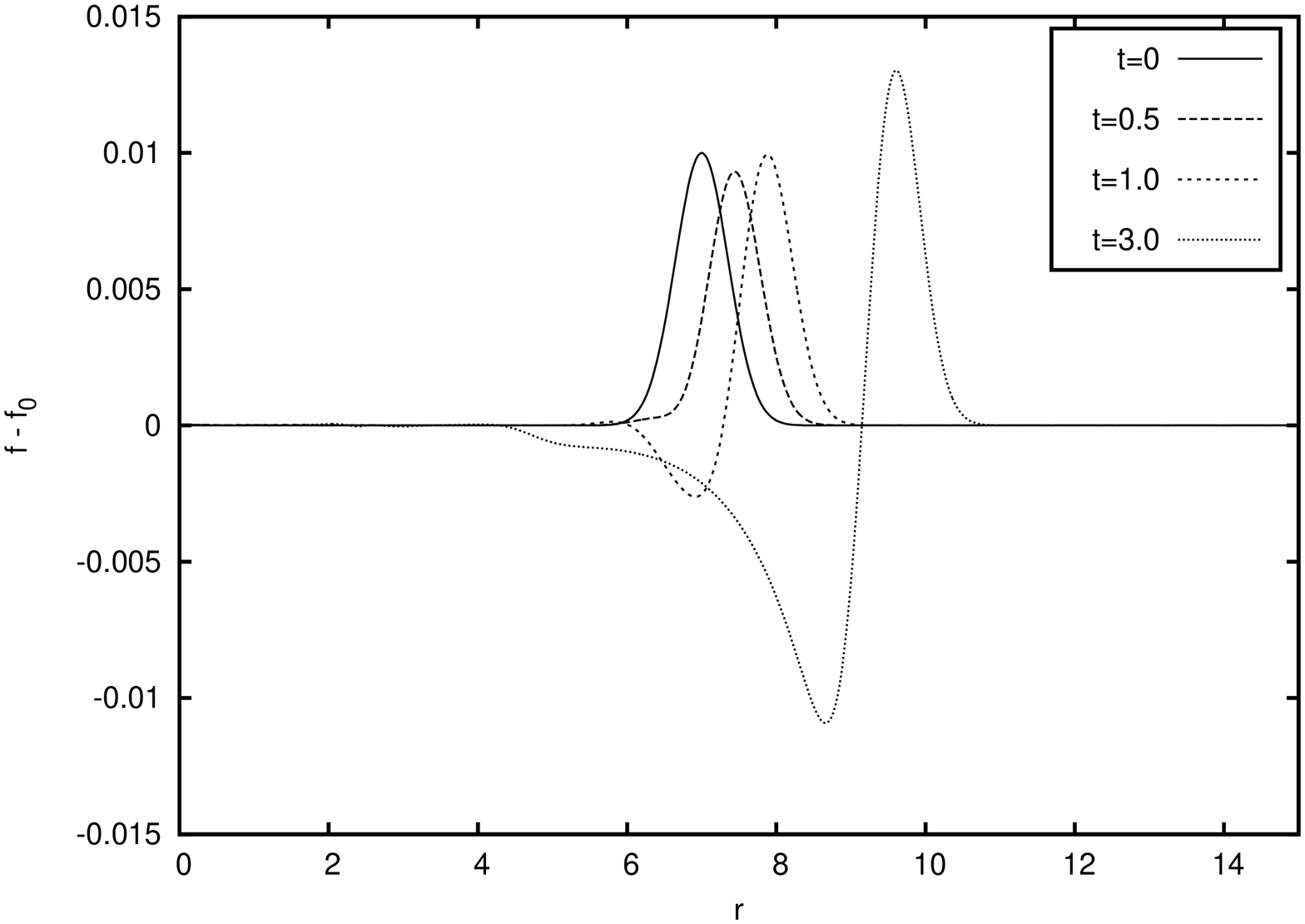}
%\vskip -13mm
\caption{Same as Figure  \ref{fig:wave1} but with the wave moving out with velocity $0.8$.}
\label{fig:wave2}
\end{figure}

Finally, we investigate  the time evolution of small perturbations (of a Gaussian form) as they  travel with different velocities towards or away from the skyrmion.  
Time snapshots of these oscillations are presented in Figures \ref{fig:wave1} and \ref{fig:wave2}.
As it can be seen  the perturbation initially travels towards the skyrmion then stops by spreading the wave packet, and finally dominates.
As the wave travels outward, its wavelength and amplitude increases, until  it is stopped due to nonlinearities.
This should be expected since the linearized field equation has no solutions with scattering asymptotics.

\begin{table}
\begin{center}
\begin{tabular}{|c|c|c|c|}
\hline
$a$  & $\omega_{\rm P}$ & $\omega$ & $T$ \\
\hline\hline
4    & 3.314         & 3.34   & 5.17\\
5    & 3.728         & 3.90   & 1.46\\
6    & 4.202         & 4.71   & 0.83\\
7    & 4.741         & 5.64   & 0.58\\
8    & 5.354         & 6.69   & 0.44\\
9    & 6.049         & 7.90   & 0.34\\
10   & 6.839         & 9.28   & 0.28\\
\hline
\end{tabular}
\end{center}
\caption{Frequencies and half-lifes of oscillations of deformed skyrmions.}
\label{tab:dyn}
\end{table}

\section{Conclusions}
By modifying the potential term of the BPS Skyrme model,  the non-compact static soliton solutions are well-defined  and sufficiently well-behaved in the whole space; while the coresponding equation  (\ref{eq:radeq})  is well-defined for sufficiently small fluctuations about a static soliton  and for sufficiently small time intervals. Also, the 
spherically symmetric sector of the dynamics of the  model has been implemented numerically and time-dependent solutions that correspond to radial pulsation and propagations of disturbances have been derived.  
The significance of the chosen potential is based on the fact that the corresponding solutions are not compactons; that is, have  analytic behaviour at the boundaries. 

 As a next step, the  investigation of  the scattering of two BPS skyrmions is going to be considered.
In this case, the initial conditions consist of   two separated  static solutions, Lorentz boosted with opposite velocities.  Due to lack of  obvious symmetries, one needs to solve  the full three-dimensional equation  numerically. We hope to report on this issue soon.

\section*{Acknowledgements}
T.I.\ thanks Christoph Adam for useful discussions.

T.I.\ acknowledges  support from  FP7, Marie Curie Actions, People, International Research Staff Exchange Scheme (IRSES-606096).
T.I.\  and \'{A}.L.\ acknowledge  support from The Hellenic Ministry of Education: Education and Lifelong Learning Affairs, and European Social Fund: NSRF 2007-2013, Aristeia (Excellence) II (TS-3647).


\end{document}